\documentclass[letterpaper, 10 pt, conference]{ieeeconf}  

\IEEEoverridecommandlockouts                              
\makeatletter
\let\NAT@parse\undefined
\makeatother

\usepackage[dvipsnames]{xcolor}

\newcommand*\linkcolours{ForestGreen}

\usepackage{times}
\usepackage{graphicx}
\usepackage{amssymb}
\usepackage{gensymb}
\usepackage{amsmath}
\usepackage{breakurl}

\usepackage{url,hyperref}
\hypersetup{
colorlinks,
linkcolor=\linkcolours,
citecolor=\linkcolours,
filecolor=\linkcolours,
urlcolor=\linkcolours}

\usepackage{algorithm}
\usepackage{algorithmic}

\usepackage[labelfont={bf},font=small]{caption}
\usepackage[none]{hyphenat}

\usepackage{mathtools, cuted}

\usepackage[noadjust, nobreak]{cite}

\usepackage{tabularx}
\usepackage{amsmath}

\usepackage{float}

\usepackage{pifont}

\newcolumntype{Y}{>{\centering\arraybackslash}X}

\usepackage[]{placeins}

\usepackage{placeins}

\usepackage{tikz}

\usepackage[framemethod=tikz]{mdframed}

\usepackage{afterpage}

\usepackage{stfloats}

\usepackage{atbegshi}
\newcommand{\handlethispage}{}
\newcommand{\discardpagesfromhere}{\let\handlethispage\AtBeginShipoutDiscard}
\newcommand{\keeppagesfromhere}{\let\handlethispage\relax}
\AtBeginShipout{\handlethispage}

\usepackage{comment}

\title{\LARGE \bf
A NOVEL METHOD FOR SCHIZOPHRENIA CLASSIFICATION USING NONLINEAR FEATURES AND NEURAL NETWORKS

}

\author{ \parbox{3 in}{\centering Hari Prasad SV{$^1$}
         \thanks{\textit{St. Stephen's College,Delhi}$^{1}$}\\
         {\tt \small hpvariyar@gmail.com}}}

\begin{document}
\maketitle

\begin{abstract}

One notable method for recording brainwaves to identify neurological problems is electroencephalography (hereafter EEG). A trained neuro physician can learn more about how the brain functions through the use of EEGs. However conventionally, EEGs are only used to examine neurological problems (Eg. Seizures). But abnormal links to neurological circuits can also exist in psychological illnesses like Schizophrenia. Hence EEGs can be an alternate source of data for detection and classification of psychological disorders. A study on the classification of EEG data obtained from healthy individuals and individuals experiencing schizophrenia is conducted. The inherent nonlinear nature of brain waves are made use for the dimensionality reduction of the data. Nonlinear parameters such as Lyapunov exponent (LE) and Hurst exponent (HE) were selected as essential features. The EEG data was obtained from the openly available EEG database of MV. Lomonosov Moscow State university. To perform Noise reduction of the data, a more recently developed Tunable Q factor based wavelet transform (TQWT) is used . Finally for the classification, the 16 channel EEG time series is converted into spatial heatmaps using the aforementioned features. A convolutional neural network (CNN) is designed and trained with the modified data format for classification.

\end{abstract}

\section{INTRODUCTION}

Schizophrenia is classified as a  psychological disorder, where an individual loses cognition and faces significant impairment in judgment and thought processes. It is characterized by hallucinations , Delusions, Inability in social interaction etc\cite{patel_2014_schizophrenia} . The aim of this paper is to evaluate schizophrenia as a neurological disorder by examining it using non linear techniques\cite{breakspear_2006_the} .  Previous work indicates the use of hybrid LSTM/CNN based networks\cite{sun_2021_ahybrid}, use of support vector machines (SVMs) \cite{bose_2016_identification} and other various types of classifiers, with varying accuracies. A comprehensive list of previous classifiers built for schizophrenia classification is documented by Sun et al \cite{sun_2021_ahybrid}. One major difference of the following work from other documented work is the novel approach for the dimensionality reduction of the data.  For convenience, This paper is divided into multiple sections, the second section describes the EEG data that is used for analysis. The third section shows how the data is converted into an image form so that it can be analysed by a Convolution Neural Network along with the different NLD features that are used to understand the data. Fourth is dedicated to discussing data augmentation and the architecture of the neural network. Last section concludes with results obtained along with discussion and remarks.

\section{Data}

\noindent Electroencephalogram data is a time series that records the electrical activity of the brain using electrodes placed in different parts of the scalp. The data acquired by these electrodes are voltage signals ranging in millivolts (mV) and showcase several characteristic brain functions. A trained neuro physician can look at this data and can understand various defects and abnormalities \cite{niedermeyers}. The data for this work is obtained from the openly accessible Schizophrenia database hosted by the MV Lomonosov Moscow State University \cite{a2019_eeg}. The files are in the form of ‘.eea’ (standard ASCII data) which includes time series numerical values obtained from the standard 10-20 placement system \cite{homan_1987_cerebral} (see figure). The two sets obtained are classified as ‘Healthy Adults ’ (39) and ‘Adults experiencing symptoms of Schizophrenia’ (45). Each datafile contains a column with 16 electrode positions. The column can be separated into sections of 16 which contains 7680 data points. A few samples of the data are shown (Fig. 2) for reference. 

\begin{figure}[H]
    \centering
    \includegraphics[width=4.5cm,height=5cm]{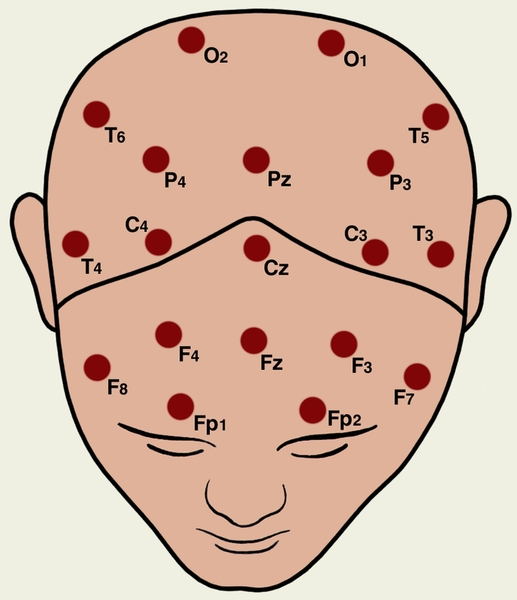}
    \caption{$10-20$ Placement system}
    \label{fig:my_label}
\end{figure}

\begin{figure}[H]
    \centering
    \includegraphics[width=8cm,height=4cm]{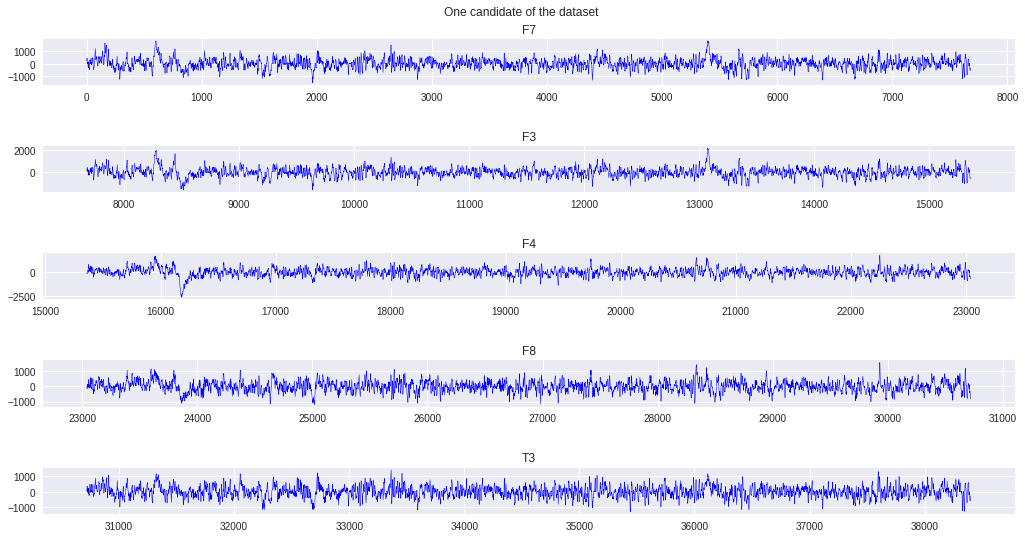}
    \caption{A sample candidate of the data (Time series from only $5$ lobes are shown.)}
    \label{fig:my_label}
\end{figure}

The obtained data is subjected to a filtering algorithm for the reduction of noise.  Due to the complexity of the time series no discernible periodicity is observed. Hence to analyse these signals, time localized wavelets are utilized thus by converting into frequency domain \cite{kumar_2017_wavelet} \cite{hamad_2016_feature}.

\subsection{Noise reduction and Wavelet Transforms}

Wavelet transform is a more robust way of performing spatio-temporal analysis on time series data. The generic idea of wavelet transform is to use a wavelet (A spatially localized wave) to analyse non stationary, transient signals \cite{elifderyabeyli_2008_waveletmixture}. Wavelet transforms are extensively used in biomedical signal analysis. In the present case, instead of using it for converting the data into frequency domain, it is used for performing noise reduction. The time series signals will eventually be subjected for the calculation of non linear features and it is essential that it is noise - free . So for this a recently developed Tunable Q wavelet transform (TQWT) developed by Selesnick \cite{selesnick_0_wavelet} is used. 

First, each channel time series is subjected to a frequency split by setting the q value as 6 and redundancy  r as 5. The series was then split into 10 energy sub bands. After the channels are done splitting, they are combined together using an inverse TQWT algorithm based on the same technique.  The entire denoising and recompilation of data was done in python by using the open source TQWT package available on github by Hajek-Lellmann.  The immediate results in the noise reduction can be seen in the heatmaps Fig. 3. All of the time series signals are subjected to the same procedure accordingly afterwards.

\begin{figure}[H]
    \centering
    \includegraphics[width=4.25cm,height=4cm]{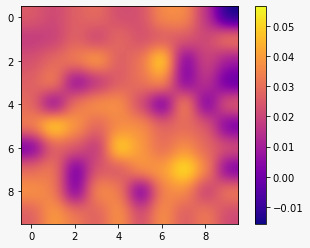}
    \includegraphics[width=4.25cm,height=4cm]{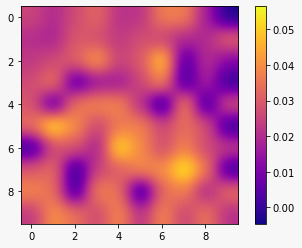}
    \caption{Heat maps before (\textit{left}) and after (\textit{right}) applying TQWT}
    \label{fig:my_label}
\end{figure}

\section{Heatmap generation and NLD features}

The idea of the Heatmap generation using non linear features in the spatial features is proposed by Kutepov, Ilya E et al.\cite{kutepovilyae_2020_eeg}. For these spatial heat maps, First the required features are calculated which are mentioned in the subsection(s) 3.1 and 3.2. After the features are calculated that of each electrode, it is mapped into a 4x4 grid. These maps represent a topological view of the scalp (Fig. 4).These heatmaps perform significant dimensionality reduction of the data thus by reducing the complexity. 

\begin{figure}[H]
    \centering
    \includegraphics[width=5cm,height=5cm]{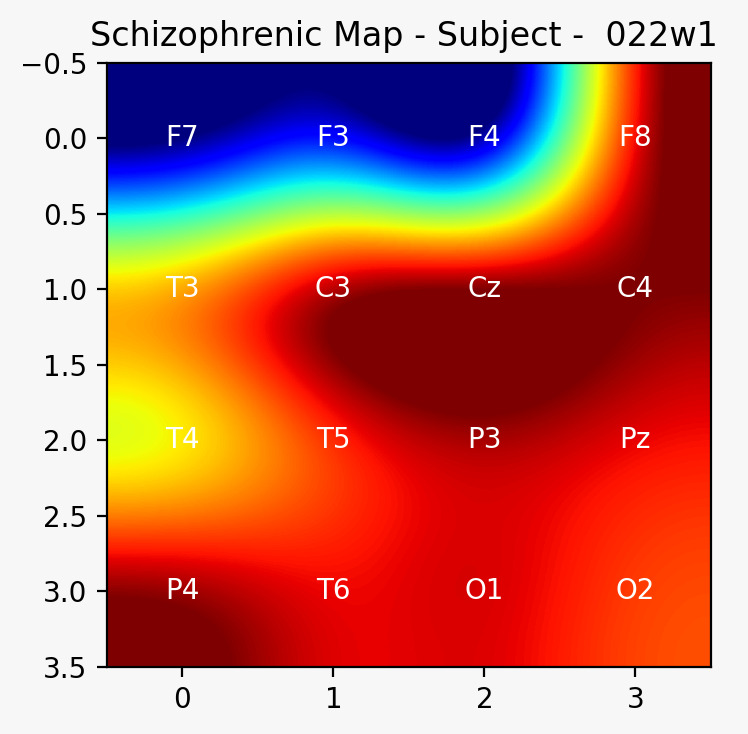}
    \caption{Electrode mapping in $4 \times 4$ grid}
    \label{fig:my_label}
\end{figure}

\subsection{Calculation of features  - Lyapunov Exponent}

Lyapunov exponent (LE) is a crucial parameter in determining the stability behaviour \cite{liapounoff_1948_probleme} of chaotic systems. LE is often regarded as the measure of divergence from the initial trajectory of the dynamical variable. The concept of LEs in dynamical systems was first proposed by Oseledec \cite{i_1968_a} as a measure of the stability of the system itself. A quantitative idea of LE can be  mathematically characterized as follows; Consider  a system with some trajectory Z(t). The divergence of the trajectory can be related to the initial trajectory  . We can define a ‘measure of divergence’ as $\lambda$ and can be related as.

\begin{equation}
    |\delta Z(t)| \approx e^{\lambda t} |\delta Z_{0}|
\end{equation}

From this. Lyapunov exponent (LE) can be defined as,

\begin{equation}
    \lambda (x_o) = \lim _{n\to\infty} \frac{1}{n}\sum _{i=0}^{n-1}\ln \left | f'(x_i) \right |
\end{equation}

For a system, there can be a spectrum of LE values according to its evolution and this spectrum LE values can be calculated (Eckmann et al. \cite{eckmann_1986_liapunov}).Here, only the maximal value of the LEs are utilized, or often termed in literature as the Maximal Lyapunov Exponent (MLE). The value of MLE is thus calculated using Rosenstein’s Algorithm proposed by Rosenstein et al. \cite{rosenstein_1993_a} using the open source python package ‘nolds’ \cite{schlzel_nolds}

\subsection{Calculation of features - Hurst Exponent}

Hurst Exponent (HE) can be defined as a measure of the long term memory of the series. The idea of the Hurst exponent was initially developed on the work done on hydrology on optimum dam requirements on the Nile river by Hurst et al\cite{hurst_1956_the}. Since the exponent generally measures the amount of autocorrelations in a time series, it gives an idea of the evolution of the series and the lag between the adjacent elements. It indicates the long term dependencies of a series, thus the measure of it is often used to determine how much the system is closely correlated to each other. The value of HE is usually between 0 and 1. A value of the exponent between 0 and 0.5 indicates the system can have high and low correlations periodically for a longer range. 0.5 denotes no correlation at all.

\noindent The Hurst exponent (HE) can be mathematically defined as,

\begin{equation}
\left ( \frac{R}{S} \right )_s = Ks^H    
\end{equation}

Where H denotes the exponent, R and S indicate the range of first ‘s’ cumulative deviations from mean and the series sum for ‘s’ entries respectively.  K is an arbitrary constant. For the calculation of HE in case of EEG time series, the previously mentioned ‘nolds’ package is utilised. After the values of Lyapunov exponent and Hurst exponent are calculated, they are used for the construction of the heat maps and fed into the classifier. 

\section{Data augmentation and CNN architecture}

After the data has been prepared, a convolutional neural network (CNN) was designed. Conventional Multilayer Perceptron feedforward (MLP) networks are not capable enough to classify multidimensional data \cite{driss_2017_a} . Hence much more robust convolutional neural networks (CNNs) are used. CNNs are particularly used in image classification (Fig. 5) or in deep neural networks \cite{driss_2017_a}. 

\subsection{Data augmentation}

To train a neural network often requires tremendous amounts of training data; for an improved classification and to prevent overfitting. The dataset used for this analysis is fairly small. To address this issue, data augmentation techniques are used. Data augmentation is a technique used when the source data from training a network is comparatively minimal \cite{amikoajczyk_0_data}. The classifier is not intended to classify the specific spatial features of the EEG from the heatmap. Elements like symmetry of the inputs can be ignored while fed into the network.  The heatmaps can be flipped horizontally along an axis as well vertically, or can be zoomed in, etc., using image processing techniques. The Tensorflow Keras \cite{abadi2016tensorflow} framework allows standard data augmentation to be performed in case of the input data. After sufficient data is generated for feeding into the network, all images are reshaped into an array of (224,224,3) to maintain uniformity along the training.

\begin{figure}[H]
    \centering
    \includegraphics[width=8cm,height=4cm]{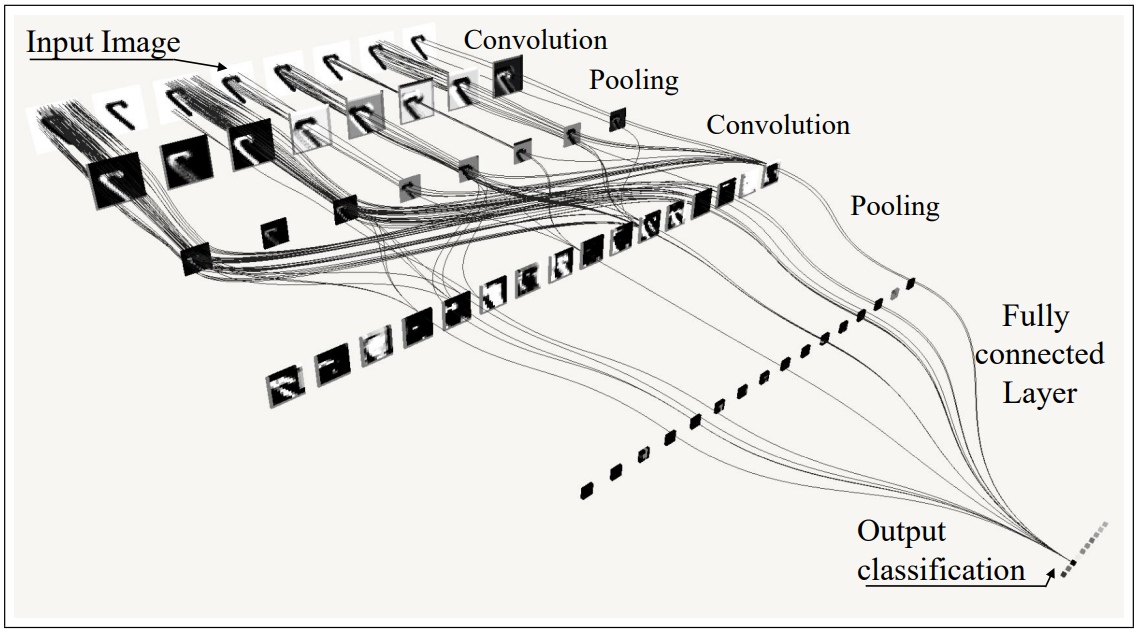}
    \caption{A conventional CNN performing image classification \cite{driss_2017_a}}
    \label{fig:my_label}
\end{figure}

\subsection{CNN architecture}

The CNN is then designed accordingly. The hyperparameters of the network were varied with some trial and error for the best results. The final CNN consisted of 3  layers with input layer taking in shapes of (224,224,3) and output layer of only one neuron. The input layer is a ‘MobilenetV2’ layer which has been previously known to perform best in performing image classification in mobile systems which have less computational efficiency \cite{sandler_2019_mobilenetv2}. The hidden layer consists of 120 neurons and ReLU activation and the output layer was activated by a sigmoid function. An adagrad optimizer, which is based on gradient descent, was used and the loss function was set to be binary crossentropic. The network was fully built and tested using Tensorflow keras framework \cite{abadi2016tensorflow}.

\section{Results and conclusions}

The accuracy of the neural network is evaluated after training, the training accuracy was 75\%. The accuracy of the network improved over the course of the training epochs. The plots (Fig.6)  shows an increase in accuracy and steady decrease in loss. 

\begin{figure}[H]
    \centering
    \includegraphics[width=7cm,height=5cm]{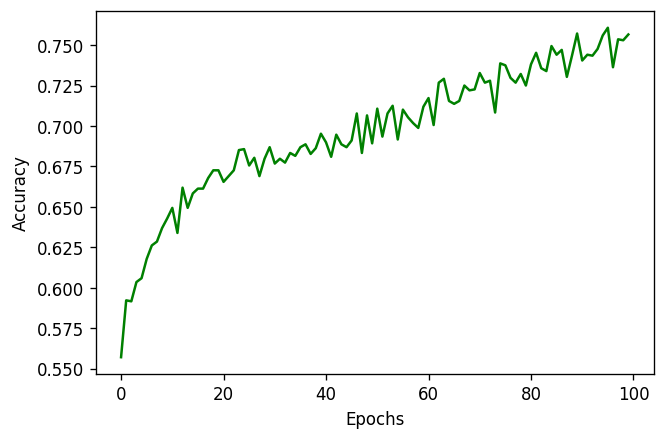}
    \caption{Accuracy}
    \label{fig:my_label}
\end{figure}

\begin{figure}[H]
    \centering
    \includegraphics[width=7cm,height=5cm]{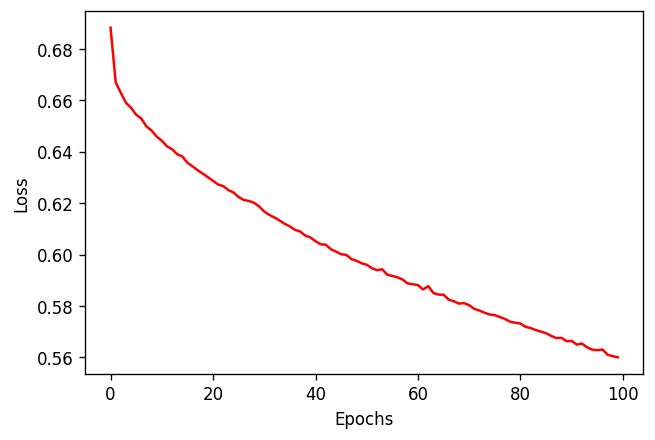}
    \caption{Loss}
    \label{fig:my_label}
\end{figure}

The results suggest the accuracy only to be 75\% which compared to that of the existing classifiers is fairly less. A low accuracy for this network was expected due to the very small training data pool. The network accuracy can be improved significantly by increasing the training data size as well as fine tuning the hyperparameters of the network. The use of other non linear features are also a possible solution to the problem. Here the results are based on the HE based heatmap data, which exhibited much more distinguishable features than that of  the Lyapunov maps. A hybrid network is also proposed which takes in multiple heatmaps as inputs which represents multiple features; Non linear as well as statistical.

\section{Acknowledgements}

\noindent The author thanks for the suggestions provided by Dr. Geetanjali Sethi, from the Department of Physics, St. Stephen’s College, Delhi. Fellow colleagues Ms. Sandra Elsa Sanjai and Mr. Vedanta Thapar (St. Stephen’s College, Delhi) are also thanked for their valuable thoughts and comments on this work.

\bibliographystyle{ieeetr}
\bibliography{bibliography}

\clearpage

\end{document}